\documentclass[nofootinbib,twocolumn,pra,letterpaper,longbibliography]{revtex4-2}
\usepackage{latexsym,amsmath,amssymb,amsfonts,graphicx,color,amsthm}
\usepackage{enumerate}
\usepackage{braket}
\usepackage{adjustbox}
\usepackage{footnote}
\usepackage[dvipsnames]{xcolor}
\usepackage{tipa}
\usepackage{textcomp}
\usepackage{fontenc}
\usepackage{mathrsfs}
\usepackage{color}
\usepackage{colortbl}
\usepackage{graphics,graphicx}
\usepackage{txfonts}

\usepackage[unicode=true,pdfusetitle, bookmarks=true,bookmarksnumbered=false,bookmarksopen=false, breaklinks=false,pdfborder={0 0 0},backref=false,colorlinks=false] {hyperref}
\hypersetup{ colorlinks,linkcolor=myurlcolor,citecolor=myurlcolor,urlcolor=myurlcolor}

\definecolor{myurlcolor}{rgb}{0,0,0.7}

\usepackage{float}
\usepackage{hyperref}

\newcommand{\cC}{\mathcal{C}}

\newcommand{\cG}{\mathcal{G}}
\newcommand{\cH}{\mathcal{H}}
\newcommand{\cI}{\mathcal{I}}
\newcommand{\cJ}{\mathcal{J}}
\newcommand{\cK}{\mathcal{K}}
\newcommand{\cL}{\mathcal{L}}

\newcommand{\cN}{\mathcal{N}}
\newcommand{\cO}{\mathcal{O}}

\newcommand{\cS}{\mathcal{S}}

\newcommand{\Id}{\mathbb{I}}
\newcommand{\tr}{\text{Tr}}

\newcommand{\comp}{\hbox{\hskip0.75mm$\circ\hskip-1.0mm\circ$\hskip0.75mm}}

\newtheorem{theorem}{Theorem}

\newtheorem{definition}{Definition}

\newtheorem{example}{Example}

\newtheorem{conjecture}{Conjecture}[theorem]

\begin{document}

\title{Information leak and incompatibility of physical context: A modified approach}
\author{Arindam Mitra$^{1,2}$}
\email{amitra@imsc.res.in}
\affiliation{$^1$Optics and Quantum Information Group, The Institute of Mathematical Sciences,
C. I. T. Campus, Taramani, Chennai 600113, India.\\
$^2$Homi Bhabha National Institute, Training School Complex, Anushakti Nagar, Mumbai 400094, India.}
\author{Gautam Sharma$^{1,2}$}
\email{gautam.oct@gmail.com}
\affiliation{$^1$Optics and Quantum Information Group, The Institute of Mathematical Sciences,
C. I. T. Campus, Taramani, Chennai 600113, India.\\
$^2$Homi Bhabha National Institute, Training School Complex, Anushakti Nagar, Mumbai 400094, India.}
\author{Sibasish Ghosh$^{1,2}$}
\email{sibasish@imsc.res.in}
\affiliation{$^1$Optics and Quantum Information Group, The Institute of Mathematical Sciences,
C. I. T. Campus, Taramani, Chennai 600113, India.\\
$^2$Homi Bhabha National Institute, Training School Complex, Anushakti Nagar, Mumbai 400094, India.}

\date{\today}

\begin{abstract}
A beautiful idea about the incompatibility of Physical Context(IPC) was introduced in [Phys. Rev. A 102, 050201(R) (2020)]. Here, a context is defined as a set of a quantum state and two sharp rank-one measurements, and the incompatibility of physical context is defined as the leakage of information while implementing those two measurements successively in that quantum state. In this work, we show the limitations in their approach. The three primary limitations are that, (i) their approach is not generalized for POVM measurements and (ii), they restrict information theoretic agents Alice, Eve and Bob to specific quantum operations and do not consider most general quantum operations i.e., quantum instruments and (iii), their measure of IPC can take negative values in specific cases in  a more general scenario which implies the limitation of their information measure. Thereby, we have introduced a generalization and modification to their approach in more general and convenient way, such that this idea is well-defined for generic measurements, without these limitations. We also present a comparison of the measure of the IPC through their and our method. Lastly, we show, how the IPC reduces in the presence of memory using our modification, which further validates our approach.
\end{abstract}

\maketitle
\section{Introduction}
Measurement incompatibility is a key feature of Quantum theory which distinguishes it from the classical world \cite{Heinosaari_2016}. A pair of observables are incompatible if they are not measurable simultaneously,i.e., their outcomes can not be obtained jointly via a single joint measurement.  Today, the connections among incompatibility, non-locality and
steering are well known \cite{Brunner__nonloc_incomp,bush_bell_steer_incomp}. Non-classical features like Bell inequality violation as well as steering can be demonstrated 
only using incompatible measurements \cite{Barnett_nonloc_incomp,Uola_steer_incomp}. It is also well known that incompatible measurements provide an advantage over compatible measurements in several information-theoretic tasks in quantum information theory \cite{qkd_d,mubs}. Measurement compatibility  can be characterized as the existence
of a common (i.e, constructed using same ancilla state and Hilbert space) commuting Naimark extensions \cite{Mitra_Naimark_incomp,PhysRevA.89.032121,BENEDUCI2017197}. It has been recently shown there are several layers of classicality inside the compatibility of measurement\cite{Heino_layers_of_class, Mitra_layers_of_class}. 

Recently, a novel idea was presented in ref.\cite{Angelo} to get a better understanding of non-classicality associated with incompatibility. The authors of \cite{Angelo} introduced the concept of incompatibility of the physical context(IPC), which is a function of a given context, where a context comprises of a quantum state and two measurements. In a way their measure of IPC captures the notion of non-classicality associated with the context, as it vanishes when the state is a maximally mixed state or the measurements are commuting with each other. It was defined as the difference between the information remaining in a quantum state after the first sharp measurement and after the second sharp measurement. Moreover, the IPC is also linked with the information leakage when an eavesdropper performs a measurement on the state being transferred in a QKD like game \cite{note}.

However, as we will show that the approach of \cite{Angelo} has several limitations. Firstly, They restrict information theoretic agents Alice, Eve and Bob to specific quantum operations and did not consider most general quantum operations i.e., quantum instruments. 	Secondly, if we do not restrict Alice, Eve and Bob to specific quantum operations which they did, then their measure of IPC can take negative value, which implies that the state after second measurement by eavesdropper Eve has more information than the state after first measurement which physically does to not make sense. Thirdly, it is not possible to extend this idea to a generic POVM measurements through their approach and without introducing quantum instruments. Fourthly, in presence of memory the IPC can increase which is against the intuition that incompatibility is non-increasing as we add memory.

In this work, we have generalized the idea of IPC for POVMs and modified the corresponding information measure. Our measure of the IPC can never be negative and it is non-increasing on addition of memory. We also demonstrate the usefulness of the modified IPC measure through a QKD like scenario as an example. In this way our approach has a wider applicability.
 
The rest of this paper is organised as follows: In section \eqref{sec:prelims}, we discuss the preliminary concepts necessary for this paper. Then, we discuss the limitations of the approach given in \cite{Angelo} and discuss our main results in section \eqref{sec:main}. Further, in section \eqref{sec:memory}, we include the presence of memory in our analysis. Finally, in section \eqref{sec:conc} we summarise our work and discuss about future direction.
\section{Preliminaries}\label{sec:prelims}
\subsection{Observables and Channels}
An observable $A$ with outcome set $\Omega_A$ in quantum mechanics, is a collection of positive hermitian  matrices $\{A(x)\mid x\in\Omega_A\}$ such that $\sum_xA(x)=\Id$. A pair of observables $(A,B)$ acting on same $d$ dimensional Hilbert space $\cH$ and with outcome sets $\Omega_A$ and $\Omega_B$ respectively, is compatible if there exist a joint observable $\cG$ acting on same Hilbert space $\cH$ and outcome set $\Omega_A\times\Omega_B$ such that for all $\rho\in\cS(\cH)$, $x\in\Omega_A$ and $y\in\Omega_B$

\begin{eqnarray}
A(x)=\sum_y\cG(x,y);
B(y)=\sum_x\cG(x,y)
\end{eqnarray}
 where $\cS(\cH)$ is the state space. Only for PVMs, compatibility implies commutativity. We denote the set of all observables as $\cO$.
 
 On the other hand, a quantum channel is a CPTP map from one state space $\cS(\cH_1)$ to another state space $\cS(\cH_2)$, where $\cH_1$ and $\cH_2$ are two Hilbert spaces. We denote the concatenation of two quantum channels $\Lambda_1$ and $\Lambda_2$ as $\Lambda_1\circ\Lambda_2$. Therefore, for all $\rho\in\cS(\cH)$ $(\Lambda_1\circ\Lambda_2)(\rho)=\Lambda_1(\Lambda_2(\rho))$. Consider two quantum channels $\Gamma:\cS(\cH_1)\rightarrow\cS(\cH_2)$ and $\Lambda:\cS(\cH_1)\rightarrow\cS(\cH^{\prime}_1)$. If there exist a quantum channel $\Theta:\cS(\cH^{\prime}_1)\rightarrow\cS(\cH_2)$ such that $\Gamma=\Theta\circ\Lambda$ holds, we denote it as $\Gamma\preceq\Lambda$. If both $\Gamma\preceq\Lambda$ and $\Lambda\preceq\Gamma$ hold, we denote it as $\Gamma\simeq\Lambda$ and we call it as $\Gamma$ and $\Lambda$ are concatenation equivalent. We denote the set of all channels concatenation equivalent to channel $\Lambda$ as $[\Lambda]$.
 
 There exists a special type of channel known as \textit{completely depolarising channel}, which we will use in following section. A channel $\Sigma$ is called completely depolarising channel if for all $T\in\cL^+(\cH)$, $\Sigma(T)=\tr(T)\eta$ for some fixed $\eta\in\cS(\cH)$, where $\cL^+(\cH)$ is  set of positive linear operators on Hilbert Space $\cH$. We denote the set of all channels as $\cC$. 
\subsection{Quantum Instruments and Measurement models}
In quantum measurements, there are two equivalent concepts, namely measurement models and quantum instruments \cite{Lahti, Heinosaari_pauli}. Measurement models are descriptions of measurement process, where as instruments are the concise version of it. Consider a measured system $S$ associated with a Hilbert space $\cH_S$ and with density matrix $\rho$ and a ancilla system associated with another Hilbert space $\cH_a$ and with density matrix $\sigma_a$. To perform a measurement on a measured system, at first a joint unitary $U$ have to be applied on the composite system where, $U$ is acting on Hilbert space $\cH_S\otimes\cH_a$. Then, a pointer observable $A^{\prime}$ with outcome set $\Omega_A^{\prime}$ have to be measured on the ancilla system. Now in this process if the observable $A$ with same outcome set as $A^{\prime}$ has, is the measured on the system $S$ then for all $x\in \Omega_A$ and $\rho \in \cH_S$ we have
\begin{equation}
\tr[\rho A(x)]=\tr[U(\rho\otimes \sigma_a)U^{\dagger}(\Id\otimes A^{\prime}(x))].
\end{equation}
The average post measurement state is given by 
\begin{equation}
\Lambda(\rho)=\tr_{\cH_a}[U(\rho\otimes \sigma_a)U^{\dagger}].
\end{equation}
Here, $\Lambda$ is a quantum channel. This measurement model is specified by the quadruple $(\cH_a,\sigma_a,U,A^{\prime})$.\\
A quantum instrument $\cI$ through which the measurement of an observable $A$ can be implemented, is a collection of CP trace non-increasing maps $\{\Phi_x\}$ such that  for all $x\in \Omega_A$ and $\rho \in \cH_S$ we have
\begin{equation}
\tr[\rho A(x)]=\tr[\Phi_x(\rho)]\label{obs_Ins}
\end{equation}
and
\begin{equation}
\sum_x\Phi_x(\rho)=\Lambda(\rho)\label{obs_chan}
\end{equation}

where $\Lambda$ is a quantum channel. We call such an instrument as $A$-compatible instrument.  If $\cI=\{\Phi_x\}$ is an $A$-compatible instrument, then another instrument $\Theta\circ\cI=\{\Theta\circ\Phi_x\}$ is also an $A$-compatible instrument \cite{Heinosaari_parent_channel}, where $(\Theta\circ\Phi_x)(\rho)=\tr[\Phi_x(\rho)]\Theta\left( \frac{\Phi_x(\rho)}{\tr[\Phi_x(\rho)]}\right)$. We denote the set of all $A$-compatible instruments as $\cJ_A$.\\
Therefore, given a measurement model $(\cH_a,\sigma_a,U,A^{\prime})$, one can associate a  quantum instrument $\cI$ such that for all $x\in \Omega_A$ and $\rho \in \cH_S$ we have

\begin{equation}
\tr[\Phi_x(\rho)]=\tr[U(\rho\otimes \sigma_a)U^{\dagger}(\Id\otimes A^{\prime}(x))].\label{eq:Measurement_model/Instrument}
\end{equation}
 Similarly, given a quantum instrument it is possible to find out a measurement model such that equation \eqref{eq:Measurement_model/Instrument} holds \cite{Ozawa}. This implies that these two concepts are equivalent.\\

\subsection{Observable-Channel compatibility}
A quantum channel $\Lambda$ is compatible with an observable $A$ if there exists a quantum instrument $\cI=\{\Phi_x\}$ such that equations \eqref{obs_Ins} and \eqref{obs_chan} together hold. Otherwise, they are incompatible. If a channel $\Lambda$ and an observable $A$ are compatible, we denote it as $\Lambda\comp A$ \cite{Heinosaari_Galois}. We call $\Lambda$ as $A$-compatible channel. \textit{It is well known that completely depolarising channels are compatible with any observable} \cite{Heinosaari_pauli}. For a quantum channel $\Lambda\in\cC$ and an observable $A\in\cO$, following sets are introduced in \cite{Heinosaari_Galois}:

\begin{align}
&\tau_{c}(\Lambda)=\{X\in\cO\mid \Lambda\comp X\};\\
&\sigma_c(A)=\{\Gamma\in\cC\mid \Gamma\comp A\}.
\end{align}
Let us now write down the following theorem which was originally proved in ref.\cite{Heinosaari_parent_channel}:

\begin{theorem}
Suppose $A\in\cO$ be an observable acting on the state space $\mathcal{S(H)})$ and $(V,\cK,\hat{A})$ be its Naimark extension, i.e, $\cK$ is a Hilbert space, $V:\mathcal{H}\rightarrow \mathcal{K}$ is an isometry and $\hat{A}=\{\hat{A}(x)\}$ is a PVM such that $V^{\dagger}\hat{A}(x)V=A(x)$ for all $x\in\Omega_A$. Then,

\begin{equation}
\sigma_c(A)=\{\Lambda\in\cC\mid \Lambda\preceq\Lambda_A\}
\end{equation}

where for any state $\rho$, $\Lambda_A(\rho)=\sum_x\hat{A}(x)V\rho V^{\dagger}\hat{A}(x)$.\label{th:parent_channel}
\end{theorem}

We call $\Lambda_A$ as parent channel of $\sigma_c(A)$ and we also call the corresponding $A$-compatible instrument $\cI_A$ a parent instrument in $\cJ_A$. Clearly, $\Lambda_A$ depends on the choice of the Naimark extension. But any two parent channels are concatenation equivalent. Therefore, we have freedom to choose it.

\subsection{Holevo Bound}\label{sec:holevo}
The Holevo bound captures the maximum classical information that can be extracted from an ensemble of quantum states\cite{Chuang}. Suppose, we have an  ensemble $\mathcal{E}$=\{$p_X(x),\rho_x$\}, and our task is to determine the classical index $x$ by doing some measurements. The density matrix operator corresponding to this ensemble has the form $\rho=\sum_xp_X(x)\rho_x$. Now, we can do a measurement $\Lambda_y$, so that the information gain after doing the measurement is given by the mutual information $I=I(X;Y)$ after the measurement, where $Y$ is the random variable corresponding to the outcome of measurement. It is known that the maximum value of this mutual information is given by the Holevo bound\cite{Chuang,Wilde}, given by
\begin{align}\label{eq:holevo}
	\chi(\mathcal{E})=S(\rho)-\sum_xp_X(x)S(\rho_x)
\end{align}
where $S(\rho)$ is the Von-Neumann entropy of the state $\rho$.
It is interesting to note that the holevo bound $\chi$ is also the mutual information of a classical-quantum state of the form $\rho_{CQ}=\sum_xp_X(x)\ket{x}\bra{x}\otimes\rho_x$. Under the action of a channel $\Lambda$ the ensemble transforms as $\mathcal{E}\rightarrow \mathcal{E}^{\prime}=\{p_X(x),\rho_x^{\prime}\}$. But we know that the mutual information is non-increasing under the action of channels\cite{Wilde}, which implies that the Holevo information is also non-increasing under the action of quantum channels, i.e.,
\begin{align}\label{eq:holevo_dec}
	\chi(\mathcal{E})\geq\chi(\mathcal{E}^{\prime}).
\end{align}

\subsection{Incompatibility of physical context}\label{sec:old_ipc}
In a recent work \cite{Angelo}, the concept of IPC was introduced which was further used to show quantum resource covariance\cite{Angelo2}. To define this idea, we need the notion of context. A context is defined as $\mathbb{C}=\{\rho,X,Y\}$, where $\rho$ is an arbitrary quantum state. Also, $X=\{X_i\}$ and $Y=\{Y_j\}$ are two observables, with $X_i$ and $Y_j$ as the respective eigen projectors. 

Other than the definition of context, we also need a game using which we define the incompatibility of a context $\mathbb{C}$. The game goes like this. Alice prepares the quantum state $\rho$, and of course it has some information content which can be quantified by using any known measure. The authors in ref.\cite{Angelo} quantify the information of $\rho$ using the following 
\begin{align}\label{eq:infAngelo}
	I(\rho)=\ln d-S(\rho),
\end{align}
where $S(\rho)$=-$Tr(\rho\ln\rho)$ is the von Neumann entropy of $\rho$ and $d$ is the dimension of the Hilbert space. This information is non-negative, i.e., $I(\rho)\geq0$, is ensured because $S(\rho)\leq\ln d$. After state preparation, Alice performs a noisy measurement with $X$ on the prepared state, so that $\rho$ transforms as
\begin{align}\label{eq:xsharp}
	\rho \rightarrow \mathcal{N}_X(\rho)=\sum_{i=1}^dX_i\rho X_i.
\end{align}
So, after this operation the information content in the state $\mathcal{N}_X(\rho)$ is $I_1=I(\mathcal{N}_X(\rho))$. This state $\mathcal{N}_X(\rho)$ is then delivered to Bob, who verifies the information content of the state. In case Bob finds that there is no loss of information, Alice and Bob will agree that the channel is free from information leakage. 

But it might happen that there is an eavesdropper, Eve, who performs a noisy measurement Y on the state $\mathcal{N}_X(\rho)$, before it is delivered to Bob. The state is then transformed as 

\begin{align}\label{eq:ysharp}
	\mathcal{N}_X(\rho) \rightarrow (\mathcal{N}_{Y}\circ\mathcal{N}_{X})(\rho)=\mathcal{N}_{YX}(\rho)=\sum_{j=1}^dY_j\mathcal{N}_{X}(\rho) Y_j.
\end{align}
Thus, the information content in the state $\cN_{XY}(\rho)$ is $I_2=I(\mathcal{N}_{YX}(\rho))$. And therefore, the leakage in the information content is given by 
\begin{align}\label{eq:old_incomp_phys_cont}
	\mathscr{I}_C=I_1-I_2&=I(\mathcal{N}_{X}(\rho))-I(\mathcal{N}_{YX}(\rho)),\nonumber \\
	&=S(\mathcal{N}_{YX}(\rho))-S(\mathcal{N}_{X}(\rho)).
\end{align}
Hence, only if $\mathscr{I}_C>0$, Alice and Bob will know that there is information leakage from the channel. Notice that $\mathscr{I}_C=0$ in two kind of scenarios: 1) If $X$ and $Y$ commute with each other and 2) if $\rho$ is a maximally mixed state. In the first scenario $\mathcal{N}_X(\rho)=\mathcal{N}_{YX}(\rho)$ because the two operators are compatible with each other. And in the second type of scenarios $I_1=0$ and there is no information to loose which results in $I_1=I_2$. Thus, we require the incompatibility $I_C$ to be non-zero for Bob to detect any leakage of information

Hence, the concept of IPC can be defined as 

\begin{definition} Context incompatibility is the resource encoded in
a context  $\mathbb{C}=\{\rho,X,Y\}$ that allows one to test the safety of a
communication channel against information leakage. It is quantified
as $\mathscr{I}_C=I_1-I_2=I(\mathcal{N}_{X}(\rho))-I(\mathcal{N}_{YX}(\rho))$. It is operationally related to the amount of information lost from the system under an external measurement.
\end{definition}
\section{Main Results}\label{sec:main}
\subsection{Limitations of incompatibility of physical context}
In this section, we discuss the limitations of approach given in \cite{Angelo}.
\begin{enumerate}
\item First of all, according to the approach given in \cite{Angelo}, the post-measurement states after measuring sharp observable $X\in\cO$ on a quantum state $\rho$ is $\cN_{X}(\rho)$. Therefore, to measure an observable $X$, Alice and Eve both are restricted to use a particular channel $\cN_{X}\in \cC$, or equivalently they are restricted to use a particular quantum instrument $\cI_{X}=\{\Phi_{X}(x)\}$ such that $\sum_{x}\Phi_{X}(x)=\cN_{X}$. Since, we have no control atleast over eavesdropper Eve, there is no reason to assume such a restriction.

\item Second, to generalize it, suppose we remove such restriction, i.e., to measure an observable, now Alice and Eve can use all possible instruments that are compatible with that observable. Then to measure the observable $X$ if Alice uses an arbitrary instrument $\cI^{\prime}_{X}=\{\Phi^{\prime}_{X}(x)\}$ such that $\Lambda^{\prime}=\sum_x\Phi^{\prime}_{X}(x)$ and to measure $Y$ Eve uses a special instrument $\cI^{depo}_{Y}=\{\Phi^{depo}_{Y}(y)\}$ such that for all $\rho\in \cS(\cH)$ and a fix pure state $\eta$, $\Lambda^{depo}_{\eta}(\rho)=\sum_y\Phi^{depo}_{Y}(y)(\rho)=\eta$ is a completely depolarising channel. Now, as $S(\eta)=0$, from equation \eqref{eq:old_incomp_phys_cont} we have

\begin{align}
\mathscr{I}_\cC&=I(\Lambda^{\prime}(\rho))-I((\Lambda^{depo}_{\eta}\circ\Lambda^{\prime})(\rho))\nonumber\\
&=-S(\Lambda^{\prime}(\rho))\nonumber \\
&\leq 0.
\end{align}

The negativity of IPC implies that the post-measurement state of Eve has more information than the post-measurement state of Alice, which does not make sense. This is because the information which Alice sends to Bob, can not be increased by the Eavesdropper. Such a problem is occurring because Von Neumann entropy is not monotonically non-increasing under action of a quantum channel. Therefore, in this general context, their information measure is not a proper information measure.

\item Thirdly, as we know that for any POVM, post measurement state depends on the quantum instrument used to implement that POVM, their results can not be generalised for POVMs without introducing quantum instruments or equivalently without introducing measurement models!
\end{enumerate}
\textit{Therefore, in our attempt  to generalize the idea of IPC for POVMs, we need to modify idea and present it in a different way which we describe in following subsections.}

\subsection{Modified measure of Information leakage }
In this section we present a generalization of the game presented in Sec.\ref{sec:old_ipc}.  Now, in the game, after the state preparation of $\rho$, instead of only doing a sharp measurement we allow Alice to perform a more generic measurement. Now, Alice performs her measurement with the POVM, $A$ on the quantum state $\rho\in\cS(\cH)$ using the $A$-compatible instrument $\cI^{\prime}_A=\{\Phi_{A,x}\}$ such that $\Lambda^{\prime}_A=\sum_x\Phi_{A,x}$ and generates the ensemble $\mathcal{E}_A=\{p_x, \rho_x\}$, where $p_x=\tr[\Phi_{A,x}(\rho)]$ and $\rho_x=\frac{\Phi_{A,x}(\rho)}{\tr[\Phi_{A,x}(\rho)]}$. Here, $\Lambda^{\prime}_A$ is the quantum channel such that $\Lambda^{\prime}_A:\cS(\cH)\rightarrow \cS(\cK)$. Furthermore, we quantify the information content of the ensemble $\mathcal{E}_A$ via the Holevo bound as:
\begin{align*}
	\chi(\rho,\cI^{\prime}_A)=S(\Lambda_A^{\prime}(\rho))-\sum_xp_xS(\rho_x).
\end{align*}

This measure of information has been previously used to quantify information gain in ref.\cite{Zhengjun}. Similarly, the eavesdropper Eve performs the POVM measurement $B$  on the quantum state $\Lambda_A^{\prime}(\rho)\in\cS(\cK)$ using the $B$-compatible instrument $\cI^{\prime}_B=\{\Phi_{B,y}\}$ such that $\Lambda^{\prime}_B=\sum_y\Phi_{B,y}$ and generates the ensemble $\mathcal{E}_B=\{p_x, \Lambda^{\prime}_B(\rho_x)\}$. Here, $\Lambda^{\prime}_B$ is the quantum channel such that $\Lambda^{\prime}_B:\cS(\cK)\rightarrow \cS(\cK^{\prime})$. It should be noted that, Alice and Bob do not have access to Eve's measurement outcomes, her measurement can be represented using a channel. Now, the information remaining in the state $(\Lambda^{\prime}_B\circ \Lambda^{\prime}_A)(\rho)$ is given by its Holevo bound, i.e.,
\begin{align*}
	\chi(\rho,\cI^{\prime}_A,\cI^{\prime}_B)=S((\Lambda^{\prime}_B\circ \Lambda^{\prime}_A)(\rho))-\sum_xp_xS(\Lambda^{\prime}_B(\rho_x)).
\end{align*}

Therefore, Bob who was expecting to receive an ensemble with information $\chi_(\rho,\cI^{\prime}_A)$, would receive a different ensemble with information content $\chi(\rho,\cI^{\prime}_A,\cI^{\prime}_B)$. Thus, the new form of information leakage of the channel is 
\begin{align}
	I^H_c(\rho,\cI^{\prime}_A,\cI^{\prime}_B)&=\chi(\rho,\cI^{\prime}_A)-\chi(\rho,\cI^{\prime}_A,\cI^{\prime}_B) \nonumber \\
	&=S(\Lambda_A^{\prime}(\rho))-S((\Lambda^{\prime}_B\circ \Lambda^{\prime}_A)(\rho))\nonumber\\
	&+\sum_xp_xS(\Lambda^{\prime}_B(\rho_x))-\sum_xp_xS(\rho_x).
\end{align}
As Holevo bound is monotonically non-increasing under the action of quantum channels, $I^H_c(\rho,\cI^{\prime}_A,\cI^{\prime}_B)\geq 0$.
When $I^H_c(\rho,\cI^{\prime}_A,\cI^{\prime}_B)>0$, Alice and Bob will be able to detect the information leakage in the channel.\\
Now, if Eve is rational, her goal will be to minimize leakage along with collecting information. Therefore, to measure $B$ she will choose an instrument such that $\cI^H_c(\rho,\cI^{\prime}_A, \cI^{\prime}_B)$ takes the minimum value. Now, let $\Lambda_B$ be a parent channel in $\sigma_c(B)$ and corresponding $B$-compatible instrument be $\cI_B$. Then, as for any other channel $\Lambda^{\prime}_B\in \sigma_B$, $\Lambda^{\prime}_B\preceq \Lambda_B$ holds and Holevo bound is monotonically decreasing under action of a quantum channel, 

\begin{align}
\chi(\rho,\cI^{\prime}_A,\cI^{\prime}_B)\leq \chi(\rho,\cI^{\prime}_A,\cI_B)~\forall\cI_B.
\end{align}
  Therefore, implementation of a parent instrument  keeps maximum amount of accessible information or equivalently maximum Holevo bound! Therefore, for a given instrument of Alice the minimum leakage of information is
 
 \begin{align}
 I^H_c(\rho,\cI^{\prime}_A,B)&=\min_{\cI^{\prime}_B}I^H_c(\rho,\cI^{\prime}_A,\cI^{\prime}_B)\nonumber\\
 &=\chi(\rho,\cI^{\prime}_A)-\max_{\cI^{\prime}_B}\chi(\rho,\cI^{\prime}_A,\cI^{\prime}_B)\nonumber\\
 &=\chi(\rho,\cI^{\prime}_A)-\chi(\rho,\cI^{\prime}_A,\cI_B)\nonumber\\
 &=I^H_c(\rho,\cI^{\prime}_A,\cI_B).
 \end{align}
Note that the choice of $B$ depends on output state space $\cS(\cK)$ of the quantum channel $\Lambda^{\prime}_A$ and in that sense, it is arbitrary.

Now, if Alice is also rational and she does not know the presence of Eve, she will try to create an ensemble with most accessible information such that the receiver i.e, Bob can get best amount of information, or equivalently, she will use an $A$-compatible instrument for which $\chi(\rho,\cI^{\prime}_A)$ is maximum. Let, $\Lambda_A$ be a parent channel in  $\sigma_c(A)$ and corresponding $A$-compatible parent instrument be $\cI_A$. Then, using arguments as above 

\begin{align}
\chi(\rho,\cI^{\prime}_A)\leq \chi(\rho,\cI_A)~\forall\cI^{\prime}_A.
\end{align}

Therefore, if Alice uses the instrument $\cI_A$, in this case the information leakage will be minimum when Alice uses a parent channel from $\sigma_c(A)$, and is given by: 
\begin{align}
I^H_c(\rho,A,B)=I^H_c(\rho,\cI_A,\cI_B).
\end{align}

Clearly, if Eve uses any other quantum instrument (e.g., dimension preserving instrument) $\cI^{\prime}_B$, then $I^H_c(\rho,\cI_A,\cI^{\prime}_B)\geq I^H_c(\rho,A,B)$. Therefore, assuming both Alice and Eve to be rational, $I^H_c(\rho,A,B)$ is the \textit{appropriate amount of information leak} when the parent instruments are used. 
 
 \subsection{Incompatibility of physical context: A modified version }
First of all we modify the notion of context so that, $\mathbb{C}=\{\rho,\mathbb{X},\mathbb{Y}\}$, where $\mathbb{X}$ and $\mathbb{Y}$ are POVM measurements acting on $\cS(\cH)$ and $\cS(\cH^{\prime})$ respectively. Since, $\mathbb{X}$ and $\mathbb{Y}$ are given, to define IPC, we restrict Alice's instrument $\cI^{\prime,\cH^{\prime}}_\mathbb{X}=\{\Phi^{\prime,\cH^{\prime}}_{\mathbb{X},x}\}$ such that $\Lambda^{\prime}_\mathbb{X}=\sum_x\Phi^{\prime,\cH^{\prime}}_{\mathbb{X},x}$ and $\Lambda^{\prime}_\mathbb{X}:\cS(\cH)\rightarrow\cS(\cH^{\prime})$. We denote the set of all such $\mathbb{X}$-compatible instruments as $\cJ^{\cH^{\prime}}_\mathbb{X}$. With this restriction also, being rational, Alice's goal will be to maximize $\chi(\rho,\cI^{\prime,\cH^{\prime}}_\mathbb{X})$. Let, for some $\cI_{\mathbb{X},max}^{\cH^{\prime}}\in\cJ^{\cH^{\prime}}_\mathbb{X}$,

\begin{equation}
	\max_{\cI_{\mathbb{X}}^{\cH^{\prime}}} \chi(\rho,\cI_{\mathbb{X}}^{\cH^{\prime}})=\chi(\rho,\cI_{\mathbb{X},max}^{\cH^{\prime}}).
\end{equation}

Therefore, similar to the previous subsection, in this case the appropriate amount of information leak is

\begin{equation}
	\mathfrak{I}(\mathbb{C})=I^H_c(\rho,\cI_{\mathbb{X},max}^{\cH^{\prime}},\cI_\mathbb{Y}).
\end{equation}
For the special case of, $\cH^{\prime}=\cH$, we have

\begin{equation}
	\mathfrak{I}(\mathbb{C})=I^H_c(\rho,\cI_{\mathbb{X},max}^{\cH},\cI_\mathbb{Y}).
\end{equation}

Therefore, we can define the generalized version of IPC as: \\
\begin{definition}
 Context incompatibility is the resource encoded in a context $\mathbb{C}=\{\rho,\mathbb{X},\mathbb{Y}\}$ that allows one to test the safety of the channel against information leakage. This resource is quantified via $\mathfrak{I}(\mathbb{C})=I^H_c(\rho,\cI_{\mathbb{X},max}^{\cH},\cI_\mathbb{Y})$, where $\cI_{\mathbb{X},max}^{\cH}$ is the $\mathbb{X}$-compatible instrument that maximizes $\chi(\rho,\cI^{\cH}_\mathbb{X})$. Operationally, it is the proper information leakage in the channel caused by an external measurement on the state. 
\end{definition}
Clearly, if Eve uses any other quantum instrument (e.g., dimension preserving instrument) $\cI^{\prime}_{\mathbb{Y}}$, then $I^H_c(\rho,\cI_{\mathbb{X},max}^{\cH},\cI^{\prime}_\mathbb{Y})\geq \mathfrak{I}(\mathbb{C})$.
Moreover, if Alice performs a sharp measurement $X=\{X_i\}$, from theorem \eqref{th:parent_channel}, choosing $V=\Id$ or equivalently choosing $\cH=\cK$ and $X_i=\hat{X_i}$ we get a parent channel $\Lambda_A=\cN(X):\cS(\cH)\rightarrow\cS(\cH)$. Let $\cI_{X}=\{\Phi_{X}\}$ be corresponding $X$-compatible parent instrument. As, implementation of a parent channel keeps maximum amount of accessible information or, equivalently maximum Holevo bound, we have $\cI_{\mathbb{X},max}^{\cH}=\cI_{X}$. Then the proper information leakage will have the following form
\begin{align}
	\mathfrak{I}(\mathbb{C})&=\chi(\rho,\cI_{X})-\chi(\rho,\cI_{X},\cI_{\mathbb{Y}}),\nonumber\\
	&=S(\mathcal{N}_X(\rho))-S((\Lambda_{\mathbb{Y}}\circ\mathcal{N}_X)(\rho))\nonumber\\
	&+\sum_xp_xS(\Lambda_{\mathbb{Y}}(\rho_x))-\sum_xp_xS(\rho_x).
\end{align}
where $\Lambda_\mathbb{Y}$ is the $\mathbb{Y}$-compatible parent channel corresponding to the $\mathbb{Y}$-compatible parent instrument $\cI_{\mathbb{Y}}$.

\subsection{Relation between two definitions}
Our generalization of the measure of IPC, gives a simplified form when we demand that both Alice and Eve perform rank-1 sharp measurements $X$ and $Y$ using parent instruments $\cI_{X}\in\cJ_{X}$ and $\cI_{Y}\in\cJ_{Y}$, where $\mathcal{N}_X\in\sigma_c(X)$ and $\mathcal{N}_Y\in\sigma_c(Y)$ are corresponding channels respectively. In this case $\mathfrak{I}(\mathbb{C})$ reads as
\begin{align}\label{eq:sharp_incompatibility}
	\mathfrak{I}(\mathbb{C})=\sum_xp_xS(\cN_{Y}(\rho_x))-\mathscr{I}_C.
\end{align}

The above equation relates our generalized measure of IPC with the measure of IPC $\mathscr{I}_C$ defined in \cite{Angelo}. To compare the two measures of the IPC, we remind the reader that $\mathscr{I}_C$ is zero when 1) $X$ and $Y$ commute or 2) $\rho$ is a maximally mixed state (see sec\ref{sec:old_ipc}). Coming to the new measure of IPC we find that $\mathfrak{I}(\mathbb{C})=0$ whenever $X$ and $Y$ commute, because then  $\cN_{Y}(\rho_x)$ are pure states. However, $\mathfrak{I}(\mathbb{C})$ is not necessarily equal to zero when $\rho$ is a maximally mixed state(as in the Eq. (\ref{eq:sharp_incompatibility}), $\mathscr{I}_C$ is zero but $S(\cN_{Y}(\rho_x))$'s are not zero). 

This implies that our measure captures the incompatibility of a context even when the state (belonging to the context) is a maximally mixed state. This is unlike the previous measure of IPC $\mathscr{I}_C$ given in ref \cite{Angelo}, which says that the context is compatible if the state is a maximally mixed state. This difference arises from the fact that the Holevo quantity(unlike the information measure in Eq.(\ref{eq:old_incomp_phys_cont}), which represents the extractable information, can be non-zero for an ensemble created from measurement on a maximally mixed state. We show the importance of the new IPC measure through the following example.
\begin{example}\label{ex:Application}
Consider a scenario in which Alice is randomly implementing $\sigma_x$ and $\sigma_z$ measurements on the maximally mixed state with equal probabilities and generates ensemble $\{\{\frac{1}{2},\ket{0}\bra{0}\},\{\frac{1}{2},\ket{1}\bra{1}\}\}$ and $\{\{\frac{1}{2},\ket{+}\bra{+}\},\{\frac{1}{2},\ket{-}\bra{-}\}\}$  respectively, for Bob. This is a QKD like situation. Now, the possible bases of  measurements are known for eavesdropper Eve. But she does not know which measurement is exactly implemented in a particular run. Therefore, she is randomly measuring $\sigma_x$ and $\sigma_z$ on the ensemble created by Alice. In this case, if we use the measure of IPC from Eq.(\ref{eq:old_incomp_phys_cont}), we get the following as the average IPC
\begin{align}
\mathscr{I}_{avg}&=\frac{1}{4}\mathscr{I}(\frac{\Id}{2}, \sigma_z, \sigma_z)+\frac{1}{4}\mathscr{I}(\frac{\Id}{2}, \sigma_z, \sigma_x)\nonumber\\
&+\frac{1}{4}\mathscr{I}(\frac{\Id}{2}, \sigma_x, \sigma_z)+\frac{1}{4}\mathscr{I}(\frac{\Id}{2}, \sigma_x, \sigma_x\nonumber)\\
&=0
\end{align}
Therefore, according to this analysis the information leakage is not detectable. But it is well established fact that if Alice and Bob declare the basis of their measurements Eve will be detected since her operation disturbs the ensemble. Therefore, this measure of IPC is not very useful here.
Instead, if we use the modified measure of IPC, we get
\begin{align}
\mathfrak{I}_{avg}&=\frac{1}{4}\mathfrak{I}(\frac{\Id}{2}, \sigma_z, \sigma_z)+\frac{1}{4}\mathfrak{I}(\frac{\Id}{2}, \sigma_z, \sigma_x)\nonumber\\
&+\frac{1}{4}\mathfrak{I}(\frac{\Id}{2}, \sigma_x, \sigma_z)+\frac{1}{4}\mathfrak{I}(\frac{\Id}{2}, \sigma_x, \sigma_x)\nonumber\\
&=\frac{1}{4}\mathfrak{I}(\frac{\Id}{2}, \sigma_z, \sigma_x)+\frac{1}{4}\mathfrak{I}(\frac{\Id}{2}, \sigma_x, \sigma_z)\nonumber\\
&=\frac{1}{2}\ln 2\nonumber \\
&\neq 0,
\end{align}
where we have used the equation \eqref{eq:sharp_incompatibility} to arrived at the last line. This non-zero value suggests that, information leakage can be detected, as expected. Hence, the modified IPC measure is useful in this scenario.
\end{example}

In this Example we have considered that Eve is using a parent instrument for her measurement. If she uses any instrument other than the parent instrument the average information leakage will be higher than $\mathfrak{I}_{avg}$.

\section{Incompatibility of physical context in the presence of memory}\label{sec:memory}
Motivated by the work in \cite{berta}, where it was shown that in the presence of memory the total uncertainty of two measurements gets reduced, we ask the question, how the IPC will change in the presence of memory? To accommodate the presence of memory, we modify our game slightly, for the scenario where we perform only rank-one projective measurements ${X}$ and ${Y}$. 

In the modified game, our initial state $\sigma_{in}$ is the subsystem of the bipartite state $\sigma_{in,M}$, where $M$ acts as the memory.  After the ${X}$ measurement on the subsystem $\sigma_{in}$, Alice produces the bipartite ensemble ${\rho}_{AM}=\sum_xp_x\rho^x_{AM}$, where $\rho^x_M=\tr_A[\rho^x_{AM}]$ acts as the memory and Bob receives the subsystem $A$ prepared in the state $\rho_A=\tr_M[\rho_{AM}]$. On this ensemble, if we use the approach in ref.\cite{Angelo}, the information content of $\rho_A$ conditioned on memory $\rho_M$ is given by
\begin{align*} 
	I^{mem}_1=\ln d-S(A|M)=\ln d-S(\rho_{AM})+S(\rho_M),
\end{align*}
where $S(A|M)=S(\rho_{AM})-S(\rho_M)$ is the conditional entropy \cite{Wilde}. After the ${Y}$ measurement by Eve on $\rho_A$, the ensemble transforms as $\sum_xp_x\rho^x_{AM}\rightarrow\sum_xp_x(\mathcal{N}_Y\otimes \mathbb{I})(\rho^x_{AM})=\sum_xp_x\rho^x_{A'M}$, so that the remaining information content of the state $\rho_{A'}$ is
\begin{align*}
	I^{mem}_2=\ln d-S(A'|M)=\ln d-S(\rho_{A'M})+S(\rho_M),
\end{align*}
where $\mathbb{I}$ is the identity channel acting on the memory. Therefore, in the presence of memory, the expression of IPC takes the following form.
\begin{align}\label{eq:old_incomp_phys_cont_mem}
	\mathscr{I}^{mem}_C&=I^{mem}_1-I^{mem}_2\nonumber \\
	&=S(\rho_{A'M})-S(\rho_{AM}).
\end{align}
To compare the IPC with and without memory, we compare Eq.(\ref{eq:old_incomp_phys_cont}) with Eq.(\ref{eq:old_incomp_phys_cont_mem}), which gives the following
\begin{align}\label{eq:oldIPCmemchange}
	\mathscr{I}_C-\mathscr{I}^{mem}_C&=[S(\rho_{A'})-S(\rho_{A'M})]-[S(\rho_{A})-S(\rho_{AM})]\nonumber \\
	&=I^{coh}(M\rangle A')-I^{coh}(M\rangle A)\leq 0.
\end{align}
Here, $I^{coh}(M\rangle A)=S(\rho_A)-S(\rho_{AM})$ is the coherent information that is non-increasing under the action of quantum channels \cite{schumacher1,schumacher2,Wilde}. This analysis tells us that the IPC is increasing in the presence of memory, which seems contrary to the intuition that memory reduces the incompatibility.

Next, we compute the IPC in the modified game with our approach. In our case, after the ${X}$ measurement, the extractable classical information from $\rho_A$ is the mutual information of the quantum-classical ensemble $\rho_{CA}=\sum_x\ket{x}_C\bra{x}\otimes\rho^x_A$ (see sec.{\ref{sec:holevo}}). However, now it is conditioned on the memory $\rho_M$. Therefore, in the presence of memory the extractable information will be the mutual information between $\rho_C=\sum_x\ket{x}_C\bra{x}$ and  $\rho_A$, conditioned on the memory $\rho_M$ via the tripartite classical-quantum state $\rho_{CAM}=\sum_xp_x\ket{x}_C\bra{x}\otimes\rho^x_{AM}$, i.e.,   
\begin{align*}
	\mathcal{X}^{mem}_1&=S(A:C|M)\\
	&=S(A|M)+S(C|M)-S(AC|M) \\
	&=S(\rho_{AM})-S(\rho_M)+S(\rho_{CM})-S(\rho_{CAM}).
\end{align*}

Here, we have simply expanded the conditional entropies to get the final form. Also, after Eve performs her measurement ${Y}$ on the subsystem $\rho_A$, the remaining mutual information between $\rho_{A'}$ and $\rho_C$ conditioned on the memory $\rho_M$, via the classical-quantum ensemble $\sum_xp_x\ket{x}_C\bra{x}\otimes\mathcal\rho^x_{A'M}$ is given by

\begin{align*}
	\mathcal{X}^{mem}_2&=S(A':C|M)\\
	&=S(A'|M)+S(C|M)-S(A'C|M) \\
	&=S(\rho_{A'M})-S(\rho_M)+S(\rho_{CM})-S(\rho_{CA'M}).
\end{align*}

Therefore the IPC, using our approach in presence of memory, takes the following form

\begin{align}\label{eq:memory_incompatibility}
	\mathfrak{I}^{mem}(\mathbb{C})&=\mathcal{X}^{mem}_1-\mathcal{X}^{mem}_2\nonumber \\
	&=S(\rho_{AM})-S(\rho_{A'M})-S(\rho_{ACM})+S(\rho_{A'CM}) \nonumber\\
	&=S(\rho_{AM})-S(\rho_{A'M})-\sum_xS(\rho^x_{AM})+\sum_xS(\rho^x_{A'M}) \nonumber \\
	&=S(\rho_{AM})-S(\rho_{A'M})+\sum_xS(\rho^x_{A'}).
\end{align}

In the above calculations we have used the fact that $\rho^x_A$ are pure states so that $\rho^x_{AM}$ and $\rho^x_{A'M}$ are bipartite product states. Now, if we compare the IPC without and with memory in from Eq.(\ref{eq:sharp_incompatibility}) and Eq.(\ref{eq:memory_incompatibility}) respectively, we have

\begin{align}\label{eq:newIPCmemchange}
	&\mathfrak{I}(\mathbb{C})-\mathfrak{I}^{mem}(\mathbb{C})\nonumber \\  &=[S(\rho_A)-S(\rho_{AM})]-[S(\rho_{A'})-S(\rho_{A'M})]\nonumber\\
	&= I^{coh}(M\rangle A)-I^{coh}(M\rangle A')\geq 0.
\end{align}
Thus, we find that using our approach, the IPC is non-increasing in the presence of memory. It follows the intuition that the presence of memory should reduce the incompatibility, as the memory can be utilized to recover the lost information. On comparing Eq.(\ref{eq:oldIPCmemchange}) with Eq.\ref{eq:newIPCmemchange}, we find that $\mathscr{I}_C-\mathscr{I}^{mem}_C=-(\mathfrak{I}(\mathbb{C})-\mathfrak{I}^{mem}(\mathbb{C}))$. This relation strongly indicates that the information leakage content $\mathscr{I}_C$ envisaged in ref.\cite{Angelo}, is not capable of fully capturing the problems we face in a typical quantum information processing scenarios. This analysis also validates our approach for quantifying the IPC.\\

\begin{example}[Comparison of incompatibilities of a physical context with two different memories]\label{ex:memory_vs_leakge}
Suppose, $\sigma_{in}=\alpha\ket{\lambda_1}\bra{\lambda_1}+\beta\ket{\lambda_2}\bra{\lambda_2}$ be a qubit state and $S_x=\{\ket{+}\bra{+},\ket{-}\bra{-}\}$ and $S_z=\{\ket{0}\bra{0},\ket{1}\bra{1}\}$ be the sharp spin measurements along $x$ and $z$ directions respectively, where $\{\ket{\lambda_1}, \ket{\lambda_2}\}$ is the eigen basis of $\sigma_{in}$. Here $0\leq\alpha,\beta\leq 1$ and $\alpha+\beta=1$. Now, we take our physical context as $\mathbb{C}_1=(\sigma_{in}, S_z, S_x)$. We will consider the following case where Alice is using memories $M$ keeping input state $\sigma_{in}$ fixed:\\
Suppose, Alice is using a qubit memory $M$ such that $\sigma_{in,M}=p\ket{\psi_{in,M}}\bra{\psi_{in,M}}+\frac{1-p}{4}\Id_{in,M}$, where $0\leq p\leq 1$ , $\ket{\psi}_{in,M}=\sqrt{\alpha^{\prime}}\ket{\lambda_1}\ket{\lambda^{\prime}_1}+\sqrt{\beta^{\prime}}\ket{\lambda_2}\ket{\lambda^{\prime}_2}$, $\Id_{AM}=\Id_{4\times 4}$, $0\leq\alpha',\beta'\leq 1$, $\alpha'+\beta'=1$,  $\{\ket{\lambda_1}^{\prime}, \ket{\lambda_2}^{\prime}\}$ is the eigen basis of $\sigma_{M}$ and $\sigma_{M}=Tr_{in}[\sigma_{in,M}]$. Alice chooses $\alpha^{\prime}$,$\beta^{\prime}$ and $p$ such that 

\begin{eqnarray}
\alpha=p\alpha^{\prime}+\frac{1-p}{2}\label{eq:alpha_alpha^prime}\\
\beta=p\beta^{\prime}+\frac{1-p}{2}\label{eq:beta_beta^prime}
\end{eqnarray}

 hold. Then, $\tr_{M}[\sigma_{in,M}]=\sigma_{in}$. For example, when $\alpha=\frac{1}{4}$ and $\beta=\frac{3}{4}$, one possible choice is $p=\frac{3}{4}$, $\alpha^{\prime}=\frac{1}{6}$ and $\beta^{\prime}=\frac{5}{6}$.  The state of the memory is $\sigma_{M}=\tr_{in}(\sigma_{in,M})=\alpha\ket{\lambda^{\prime}_1}\bra{\lambda^{\prime}_1}+\beta\ket{\lambda^{\prime}_2}\bra{\lambda^{\prime}_2}$. Let, $q_{xy}={\braket{x|\lambda_y}}$ where $x\in\{0,1,+,-\}$ and $y\in\{1,2\}$. The bipartite ensemble, created by the $S_z$ measurement of Alice, is $\{p^{\prime}_i,\sigma^i_{AM}\}$ where, $p^{\prime}_i=\tr[(\ket{i}\bra{i}\otimes\Id)\sigma_{in,M}]= p[\alpha^{\prime}\mid q_{i1}\mid^2+\beta^{\prime}\mid q_{i2}\mid^2]+\frac{1-p}{2}$ and $\sigma^i_{AM}=\frac{(\ket{i}\bra{i}\otimes\Id)\sigma_{in,M}(\ket{i}\bra{i}\otimes\Id)}{\tr[(\ket{i}\bra{i}\otimes\Id)\sigma_{in,M}]}$ and $i\in\{0,1\}$. Now it can be easily checked that $\sigma^i_{AM}=\ket{i}\bra{i}\otimes[p\ket{\phi^{\prime}_i}\bra{\phi^{\prime}_i}+\frac{1-p}{2p^{\prime}_i}\frac{\Id}{2}]$ where $\ket{\phi_i}=\frac{1}{\sqrt{p_i}}(\sqrt{\alpha^{\prime}}q_{i1}\ket{\lambda_1}+\sqrt{\beta^{\prime}}q_{i2}\ket{\lambda_2})$.
The post-measurement average bipartite state is $\sigma_{AM}=\sum_ip^{\prime}_i\sigma^i_{AM}$. Clearly, $\sigma_A=\tr_M\sigma_{AM}=p\sum_ip^{\prime}_i\ket{i}\bra{i}+(1-p)\frac{\Id}{2}$. After, Eve's $S_x$  measurement on $A$ part, the average bipartite state will become $\sigma_{A^{\prime}M}=\frac{\Id}{2}\otimes\sigma_{M}$ where $\sigma_{M}=[p\sum_ip_i\ket{\phi^{\prime}_i}\bra{\phi^{\prime}_i}+(1-p)\frac{\Id}{2}]=\sigma_{M}$ and the average state of $A$ part becomes $\sigma_{A^{\prime}}=\frac{\Id}{2}$. 
 So, the reduction in information leak is given as
 \begin{align}
&\mathfrak{I}(\mathbb{C}_1)-\mathfrak{I}^{M}(\mathbb{C}_1)\nonumber \\ 
&=[S(\mathcal{\sigma_A})-S(\sigma_{AM})]
\hspace{0.2cm}-[S(\sigma^{\prime}_A)-S(\sigma_{A^{\prime}M})]\nonumber\\
&=[S(\mathcal{\sigma_A})-S(\sigma_{AM})]
\hspace{0.2cm}-[S(\sigma^{\prime}_A)-S(\sigma_{A^{\prime}})-S(\sigma_{M})]\nonumber\\
&=S(\sigma_{M})+S(\sigma_{A})-S(\sigma_{AM})=I(A:M)_{\sigma_{AM}}.
\end{align}
%It is easy to check that for $p=1$, the leak difference $\mathfrak{I}(\mathbb{C}_1)-\mathfrak{I}^{M}(\mathbb{C}_1)=S(\sigma_{M})$.
Now, consider a special case where $\ket{\lambda_1},\ket{\lambda_2}$ are the eigen basis of $\sigma_y$, $\alpha=\frac{1}{4}$ and $\beta=\frac{3}{4}$. In this case, $\mid q_{ij}\mid^2=\frac{1}{2}$ $\forall i\in\{0,1\}$ and $\forall j\in\{1,2\}$. Also, from equation \eqref{eq:alpha_alpha^prime} we get $\alpha^{\prime}=\frac{2p-1}{4p}$. Clearly, $\alpha^{\prime}\geq 0$ only for $p\geq \frac{1}{2}$. We plot the leakage difference $\mathfrak{I}(\mathbb{C}_1)-\mathfrak{I}^{M}(\mathbb{C}_1)$ with respect to $p$ in Fig. \eqref{fig:Plot}.
\begin{figure}[hbt!]
\includegraphics[width=8.4cm,height=6cm]{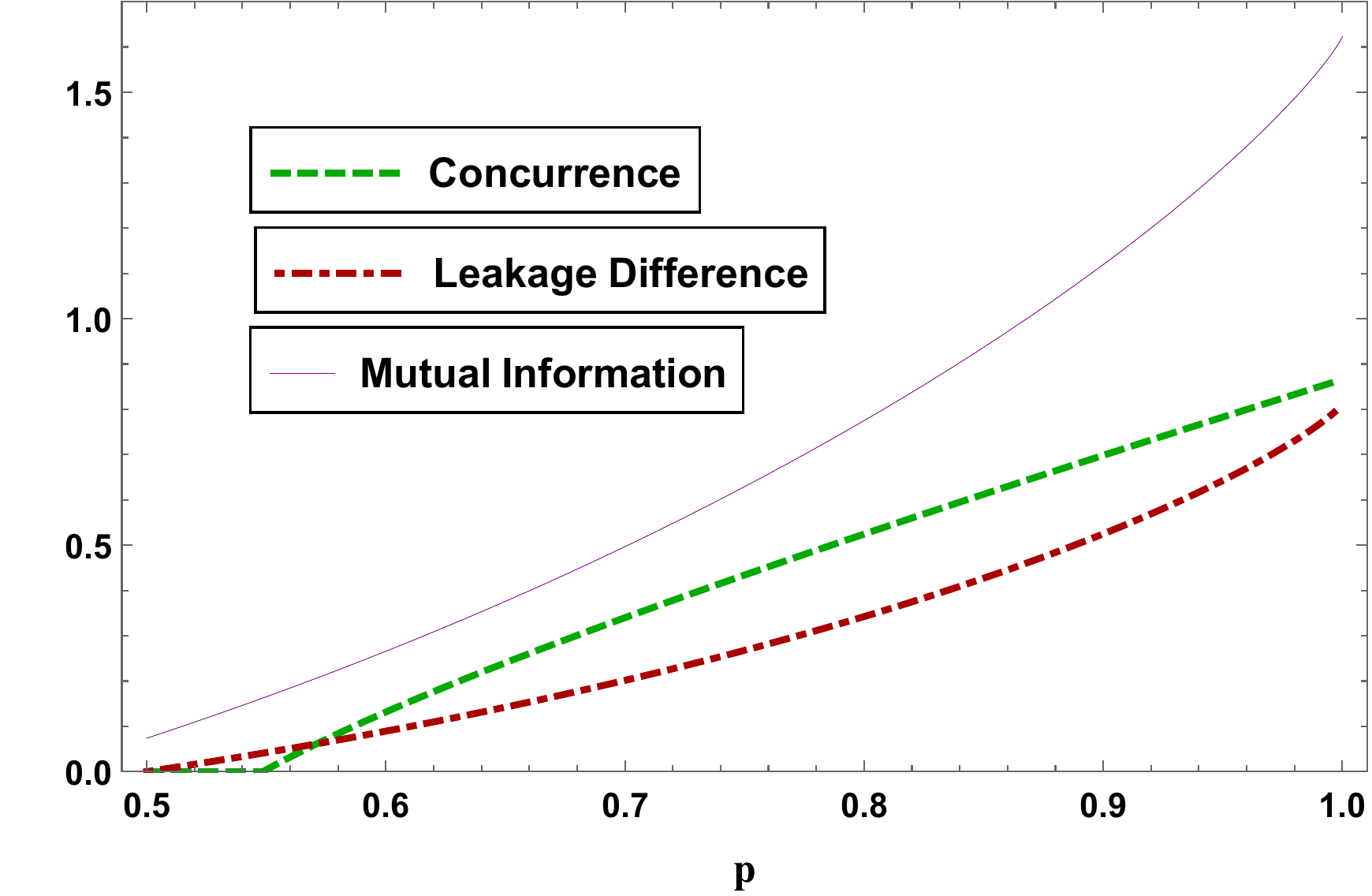}
\caption{(Colour online) Plot of Concurrence and mutual information of $\sigma_{in,M}$ and the leakage difference vs the parameter $p$. It can be seen that  Concurrence and mutual information of $\sigma_{in,M}$ and the leakage difference is monotonically increasing with respect to the parameter $p$. All quantities are normalized i.e., all of them have been divided by their maximum values.}\label{fig:Plot}
\end{figure}
To quantify the amount of memory we use the concurrence measure(ref. \cite{Wootters}) and the mutual information of the initial bipartite state $\sigma_{in,M}$. From Fig. \eqref{fig:Plot} we get that with increment of $p$, concurrence and mutual information of $\sigma_{in,M}$ and the leakage difference, are monotonically increasing with $p$. We can also say that the information leakage difference is a monotonically increasing function of both concurrence and mutual information in the state $\sigma_{in,M}$. Equivalently, we can say that the leakage with memory is monotonically decreasing with increasing value concurrence and mutual information. It can be observed from Fig.\eqref{fig:Plot}, the leakage difference is non-zero for the region $p\lessapprox 0.548$ where the concurrence is vanishing. In this region the non-zero leakage difference can be attributed to the non-vanishing mutual information.
\end{example}
The example \eqref{ex:memory_vs_leakge} suggests us to write down the following conjecture:
\begin{conjecture}
With increment of correlation between the memory and the input state, information leakage monotonically decreases. 
\end{conjecture} 
Therefore, we conclude based on the validity of the conjecture, that the presence of more memory correlation helps in reducing the leakage. 

\section{conclusion}\label{sec:conc}
In this work, we have derived the measure of an appropriate information leakage in all QKD like games. Moreover, introducing quantum instruments, we have generalised the notion of IPC for POVMs.  We  have shown the relation between previous and our approaches for sharp measurements. Our approach always leads to a non-negative measure of IPC. We have also shown that the modified IPC measure is more useful compared to the earlier IPC measure in Eq.(\ref{eq:old_incomp_phys_cont}), in a QKD like scenario as an example. Also, on including memory, our measure of IPC can never increase. In fact in example\eqref{ex:memory_vs_leakge}, we have shown that information leakage monotonically decreases with increment of correlation between input state and memory.  Thus, we have successfully modified the notion of IPC for generic measurements.\\
 
Our work opens up several future directions. Firstly, it would be useful to construct the resource theory of IPC using our measure. Further, our measure can be a useful tool for generic information-theoretic tasks, that involve transmission of classical information over quantum channels. We would like to explore how our generalized version of IPC can be related to incompatibility of POVMs.

\bibliographystyle{apsrev4-1}
\bibliography{physical_context}
\end{document}